\documentclass[10pt,a4paper,onecolumn]{article}
\usepackage{marginnote}
\usepackage{graphicx}
\usepackage{xcolor}
\usepackage{authblk,etoolbox}
\usepackage{titlesec}
\usepackage{calc}
\usepackage{tikz}
\usepackage{hyperref}
\hypersetup{colorlinks,breaklinks=true,
            urlcolor=[rgb]{0.0, 0.5, 1.0},
            linkcolor=[rgb]{0.0, 0.5, 1.0}}
\usepackage{caption}
\usepackage{tcolorbox}
\usepackage{amssymb,amsmath,dsfont}
\usepackage{ifxetex,ifluatex}
\usepackage{seqsplit}
\usepackage{xstring}

\usepackage{float}
\let\origfigure\figure
\let\endorigfigure\endfigure
\renewenvironment{figure}[1][2] {
    \expandafter\origfigure\expandafter[H]
} {
    \endorigfigure
}

\usepackage{fixltx2e} 
\usepackage[
  backend=biber,
]{biblatex}
\bibliography{paper.bib}

\let\textttOrig=\texttt
\def\texttt#1{\expandafter\textttOrig{\seqsplit{#1}}}
\renewcommand{\seqinsert}{\ifmmode
  \allowbreak
  \else\penalty6000\hspace{0pt plus 0.02em}\fi}

\makeatletter
\let\href@Orig=\href
\def\href@Urllike#1#2{\href@Orig{#1}{\begingroup
    \def\Url@String{#2}\Url@FormatString
    \endgroup}}
\def\href@Notdoi#1#2{\def\tempa{#1}\def\tempb{#2}%
  \ifx\tempa\tempb\relax\href@Urllike{#1}{#2}\else
  \href@Orig{#1}{#2}\fi}
\def\href#1#2{%
  \IfBeginWith{#1}{https://doi.org}%
  {\href@Urllike{#1}{#2}}{\href@Notdoi{#1}{#2}}}
\makeatother

\newlength{\cslhangindent}
\newlength{\csllabelwidth}
\newenvironment{CSLReferences}[3] 
 {
  \setlength{\parindent}{0pt}
  \ifodd #1 \everypar{\setlength{\hangindent}{\cslhangindent}}\ignorespaces\fi
  \ifnum #2 > 0
  \setlength{\parskip}{#2\baselineskip}
  \fi
 }%
 {}
\usepackage{calc}

\usepackage[top=3.5cm, bottom=3cm, right=2.54cm, left=2.54cm,
            headheight=2.2cm]{geometry}

\titleformat{\section}
  {\normalfont\sffamily\Large\bfseries}
  {}{0pt}{}
\titleformat{\subsection}
  {\normalfont\sffamily\large\bfseries}
  {}{0pt}{}
\titleformat{\subsubsection}
  {\normalfont\sffamily\bfseries}
  {}{0pt}{}
\titleformat*{\paragraph}
  {\sffamily\normalsize}

\usepackage{fancyhdr}
\makeatletter
\let\ps@plain\ps@fancy
\fancyheadoffset[L]{4.5cm}
\fancyfootoffset[L]{4.5cm}

\definecolor{linky}{rgb}{0.0, 0.5, 1.0}

\newtcolorbox{repobox}
   {colback=red, colframe=red!75!black,
     boxrule=0.5pt, arc=2pt, left=6pt, right=6pt, top=3pt, bottom=3pt}

\patchcmd{\@maketitle}{center}{flushleft}{}{}
\patchcmd{\@maketitle}{center}{flushleft}{}{}
\patchcmd{\@maketitle}{\LARGE}{\LARGE\sffamily}{}{}
\def\maketitle{{%
  
  \AB@maketitle}}
\makeatletter
\renewcommand\AB@affilsepx{ \protect\Affilfont}
\renewcommand\AB@affilnote[1]{{\bfseries #1}\hspace{3pt}}
\renewcommand{\affil}[2][]%
   {\newaffiltrue\let\AB@blk@and\AB@pand
      \if\relax#1\relax\def\AB@note{\AB@thenote}\else\def\AB@note{#1}%
        \setcounter{Maxaffil}{0}\fi
        \begingroup
        \let\href=\href@Orig
        \let\texttt=\textttOrig
        \let\protect\@unexpandable@protect
        \def\thanks{\protect\thanks}\def\footnote{\protect\footnote}%
        \@temptokena=\expandafter{\AB@authors}%
        {\def\\{\protect\\\protect\Affilfont}\xdef\AB@temp{#2}}%
         \xdef\AB@authors{\the\@temptokena\AB@las\AB@au@str
         \protect\\[\affilsep]\protect\Affilfont\AB@temp}%
         \gdef\AB@las{}\gdef\AB@au@str{}%
        {\def\\{, \ignorespaces}\xdef\AB@temp{#2}}%
        \@temptokena=\expandafter{\AB@affillist}%
        \xdef\AB@affillist{\the\@temptokena \AB@affilsep
          \AB@affilnote{\AB@note}\protect\Affilfont\AB@temp}%
      \endgroup
       \let\AB@affilsep\AB@affilsepx
}
\makeatother

\renewcommand\Affilfont{\sffamily\small\mdseries}
\ifnum 0\ifxetex 1\fi\ifluatex 1\fi=0 
  \usepackage[T1]{fontenc}
  \usepackage[utf8]{inputenc}

\else 
  \ifxetex
    \usepackage{mathspec}
    \usepackage{fontspec}

  \else
    \usepackage{fontspec}
  \fi
  \defaultfontfeatures{Ligatures=TeX,Scale=MatchLowercase}

\fi
\IfFileExists{upquote.sty}{\usepackage{upquote}}{}
\IfFileExists{microtype.sty}{%
\usepackage{microtype}
\UseMicrotypeSet[protrusion]{basicmath} 
}{}

\usepackage{hyperref}
\hypersetup{unicode=true,
            pdftitle={Re-Envisioning Numerical Information Field Theory (NIFTy.re): A Library for Gaussian Processes and Variational Inference},
            pdfborder={0 0 0},
            breaklinks=true}
\usepackage{color}
\usepackage{fancyvrb}

\DefineVerbatimEnvironment{Highlighting}{Verbatim}{commandchars=\\\{\}}
\newenvironment{Shaded}{}{}
\newcommand{\AlertTok}[1]{\textcolor[rgb]{1.00,0.00,0.00}{\textbf{#1}}}

\newcommand{\BuiltInTok}[1]{#1}

\newcommand{\CommentTok}[1]{\textcolor[rgb]{0.38,0.63,0.69}{\textit{#1}}}

\newcommand{\ControlFlowTok}[1]{\textcolor[rgb]{0.00,0.44,0.13}{\textbf{#1}}}

\newcommand{\DecValTok}[1]{\textcolor[rgb]{0.25,0.63,0.44}{#1}}

\newcommand{\FloatTok}[1]{\textcolor[rgb]{0.25,0.63,0.44}{#1}}
\newcommand{\FunctionTok}[1]{\textcolor[rgb]{0.02,0.16,0.49}{#1}}
\newcommand{\ImportTok}[1]{#1}

\newcommand{\KeywordTok}[1]{\textcolor[rgb]{0.00,0.44,0.13}{\textbf{#1}}}
\newcommand{\NormalTok}[1]{#1}
\newcommand{\OperatorTok}[1]{\textcolor[rgb]{0.40,0.40,0.40}{#1}}

\newcommand{\SpecialCharTok}[1]{\textcolor[rgb]{0.25,0.44,0.63}{#1}}
\newcommand{\SpecialStringTok}[1]{\textcolor[rgb]{0.73,0.40,0.53}{#1}}
\newcommand{\StringTok}[1]{\textcolor[rgb]{0.25,0.44,0.63}{#1}}
\newcommand{\VariableTok}[1]{\textcolor[rgb]{0.10,0.09,0.49}{#1}}

\let\addcontentslineOrig=\addcontentsline
\def\addcontentsline#1#2#3{\bgroup
  \let\texttt=\textttOrig\addcontentslineOrig{#1}{#2}{#3}\egroup}
\let\markbothOrig\markboth
\def\markboth#1#2{\bgroup
  \let\texttt=\textttOrig\markbothOrig{#1}{#2}\egroup}
\let\markrightOrig\markright
\def\markright#1{\bgroup
  \let\texttt=\textttOrig\markrightOrig{#1}\egroup}

\usepackage{graphicx,grffile}
\makeatletter
\def\maxwidth{\ifdim\Gin@nat@width>\linewidth\linewidth\else\Gin@nat@width\fi}
\def\maxheight{\ifdim\Gin@nat@height>\textheight\textheight\else\Gin@nat@height\fi}
\makeatother
\setkeys{Gin}{width=\maxwidth,height=\maxheight,keepaspectratio}
\IfFileExists{parskip.sty}{%
\usepackage{parskip}
}{
\setlength{\parindent}{0pt}
\setlength{\parskip}{6pt plus 2pt minus 1pt}
}
\ifx\paragraph\undefined\else
\let\oldparagraph\paragraph
\renewcommand{\paragraph}[1]{\oldparagraph{#1}\mbox{}}
\fi
\ifx\subparagraph\undefined\else
\let\oldsubparagraph\subparagraph
\renewcommand{\subparagraph}[1]{\oldsubparagraph{#1}\mbox{}}
\fi

\title{Re-Envisioning Numerical Information Field Theory (NIFTy.re): A
Library for Gaussian Processes and Variational Inference}

        \author[1, 2, 3]{Gordian Edenhofer}
          \author[1]{Philipp Frank}
          \author[1, 2, 4]{Jakob Roth}
          \author[5]{Reimar H. Leike}
          \author[2]{Massin Guerdi}
          \author[2, 6, 7]{Lukas I. Scheel-Platz}
          \author[1, 2]{Matteo Guardiani}
          \author[1, 2]{Vincent Eberle}
          \author[1, 2]{Margret Westerkamp}
          \author[1, 2]{Torsten A. Enßlin}

      \affil[1]{Max Planck Institute for Astrophysics,
Karl-Schwarzschild-Straße 1, 85748 Garching bei München, Germany}
      \affil[2]{Ludwig Maximilian University of Munich,
Geschwister-Scholl-Platz 1, 80539 München, Germany}
      \affil[3]{Department of Astrophysics, University of Vienna,
Türkenschanzstraße 17, A-1180 Vienna, Austria}
      \affil[4]{School of Computation, Information and Technology,
Technical University of Munich, Arcisstr. 21, 80333 München, Germany}
      \affil[5]{Independent Researcher, USA}
      \affil[6]{Helmholtz Zentrum München, Ingolstädter Landstraße 1,
85764 Neuherberg, Germany}
      \affil[7]{School of Medicine and Health, Technical University of
Munich, Ismaninger Str. 22, 81675 München, Germany}
  \date{\vspace{-7ex}}

\begin{document}
\maketitle

\vspace{1em}
\hypertarget{summary}{%
\section{Summary}\label{summary}}

Imaging is the process of transforming noisy, incomplete data into a
space that humans can interpret. \texttt{NIFTy} is a Bayesian framework
for imaging and has already successfully been applied to many fields in
astrophysics. Previous design decisions held the performance and the
development of methods in \texttt{NIFTy} back. We present a rewrite of
\texttt{NIFTy}, coined \texttt{NIFTy.re}, which reworks the modeling
principle, extends the inference strategies, and outsources much of the
heavy lifting to JAX. The rewrite dramatically accelerates models
written in \texttt{NIFTy}, lays the foundation for new types of
inference machineries, improves maintainability, and enables
interoperability between \texttt{NIFTy} and the JAX machine learning
ecosystem.

\hypertarget{statement-of-need}{%
\section{Statement of Need}\label{statement-of-need}}

Imaging commonly involves millions to billions of pixels. Each pixel
usually corresponds to one or more correlated degrees of freedom in the
model space. Modeling this many degrees of freedom is computationally
demanding. However, imaging is not only computationally demanding but
also statistically challenging. The noise in the data requires a
statistical treatment and needs to be accurately propagated from the
data to the uncertainties in the final image. To do this, we require an
inference machinery that not only handles extremely high-dimensional
spaces, but one that does so in a statistically rigorous way.

\texttt{NIFTy} is a Bayesian imaging library (Arras, Baltac, et al.,
2019; Selig et al., 2013; Steininger et al., 2019). It is designed to
infer the million- to billion-dimensional posterior distribution in the
image space from noisy input data. At the core of \texttt{NIFTy} lies a
set of powerful Gaussian Process (GP) models and accurate Variational
Inference (VI) algorithms.

\texttt{NIFTy.re} is a rewrite of \texttt{NIFTy} in JAX (Bradbury et
al., 2018) with all relevant previous GP models, new, more flexible GP
models, and a more flexible machinery for approximating posterior
distributions. Being written in JAX, \texttt{NIFTy.re} effortlessly runs
on accelerator hardware such as the GPU and TPU, vectorizes models
whenever possible, and just-in-time compiles code for additional
performance. \texttt{NIFTy.re} switches from a home-grown automatic
differentiation engine that was used in \texttt{NIFTy} to JAX's
automatic differentiation engine. This lays the foundation for new types
of inference machineries that make use of the higher order derivatives
provided by JAX. Through these changes, we envision to harness
significant gains in maintainability of \texttt{NIFTy.re} compared to
\texttt{NIFTy} and a faster development cycle for new features.

We expect \texttt{NIFTy.re} to be highly useful for many imaging
applications and envision many applications within and outside of
astrophysics (Arras et al., 2022; Arras, Frank, et al., 2019; Eberle et
al., 2023, 2022; Frank et al., 2017; S. Hutschenreuter et al., 2022;
Sebastian Hutschenreuter et al., 2023; Leike et al., 2020; Leike \&
Enßlin, 2019; Mertsch \& Phan, 2023; J. Roth et al., 2023; Jakob Roth et
al., 2023; Scheel-Platz et al., 2023; Tsouros et al., 2024; Welling et
al., 2021; Westerkamp et al., 2023). \texttt{NIFTy.re} has already been
successfully used in two galactic tomography publications (Edenhofer et
al., 2024; Leike et al., 2022). A very early version of
\texttt{NIFTy.re} enabled a 100-billion-dimensional reconstruction using
a maximum posterior inference. In a newer publication, \texttt{NIFTy.re}
was used to infer a 500-million-dimensional posterior distribution using
VI (Knollmüller \& Enßlin, 2019). The latter publication extensively
used \texttt{NIFTy.re}'s GPU support to reduce the runtime by two orders
of magnitude compared to the CPU. With \texttt{NIFTy.re} bridging ideas
from \texttt{NIFTy} to JAX, we envision many new possibilities for
inferring classical machine learning models with \texttt{NIFTy}'s
inference methods and a plethora of opportunities to use
\texttt{NIFTy}-components such as the GP models in classical neural
network frameworks.

\texttt{NIFTy.re} competes with other GP libraries as well as with
probabilistic programming languages and frameworks. Compared to GPyTorch
(Hensman et al., 2015), GPflow (De G. Matthews et al., 2017), george
(Ambikasaran et al., 2015), or TinyGP (Foreman-Mackey et al., 2024),
\texttt{NIFTy.re} focuses on GP models for structured spaces and does
not assume the posterior to be analytically accessible. Instead,
\texttt{NIFTy.re} tries to approximate the true posterior using VI.
Compared to classical probabilistic programming languages such as Stan
(Carpenter et al., 2017) and frameworks such as Pyro (Bingham et al.,
2019), NumPyro (Phan et al., 2019), pyMC3 (Salvatier et al., 2016),
emcee (Foreman-Mackey et al., 2013), dynesty (Koposov et al., 2023;
Speagle, 2020), or BlackJAX (Cabezas \& Louf, 2023), \texttt{NIFTy.re}
focuses on inference in extremely high-dimensional spaces.
\texttt{NIFTy.re} exploits the structure of probabilistic models in its
VI techniques (Frank et al., 2021). With \texttt{NIFTy.re}, the GP
models and the VI machinery are now fully accessible in the JAX
ecosystem and \texttt{NIFTy.re} components interact seamlessly with
other JAX packages such as BlackJAX and JAXopt/Optax (Blondel et al.,
2022; DeepMind et al., 2020).

\hypertarget{core-components}{%
\section{Core Components}\label{core-components}}

\texttt{NIFTy.re} brings tried and tested structured GP models and VI
algorithms to JAX. GP models are highly useful for imaging problems, and
VI algorithms are essential to probe high-dimensional posteriors, which
are often encountered in imaging problems. \texttt{NIFTy.re} infers the
parameters of interest from noisy data via a stochastic mapping that
goes in the opposite direction, from the parameters of interest to the
data.

\texttt{NIFTy} and \texttt{NIFTy.re} build up hierarchical models for
the posterior inference. The log-posterior function reads
\(\ln{p(\theta|d)} := \ell(d, f(\theta)) + \ln{p}(\theta) + \mathrm{const}\)
with log-likelihood \(\ell\), forward model \(f\) mapping the parameters
of interest \(\theta\) to the data space, and log-prior
\(\ln{p(\theta)}\). The goal of the inference is to draw samples from
the posterior \(p(\theta|d)\).

What is considered part of the likelihood versus part of the prior is
ill-defined. Without loss of generality, \texttt{NIFTy} and
\texttt{NIFTy.re} re-formulate models such that the prior is always
standard Gaussian. They implicitly define a mapping from a new latent
space with a priori standard Gaussian parameters \(\xi\) to the
parameters of interest \(\theta\). The mapping \(\theta(\xi)\) is
incorporated into the forward model \(f(\theta(\xi))\) in such a way
that all relevant details of the prior model are encoded in the forward
model. This choice of re-parameterization (Rezende \& Mohamed, 2015) is
called standardization. It is often carried out implicitly in the
background without user input.

\hypertarget{gaussian-processes}{%
\subsection{Gaussian Processes}\label{gaussian-processes}}

One standard tool from the \texttt{NIFTy.re} toolbox is the so-called
correlated field GP model from \texttt{NIFTy}. This model relies on the
harmonic domain being easily accessible. For example, for pixels spaced
on a regular Cartesian grid, the natural choice to represent a
stationary kernel is the Fourier domain. In the generative picture, a
realization \(s\) drawn from a GP then reads
\(s = \mathrm{FT} \circ \sqrt{P} \circ \xi\) with \(\mathrm{FT}\) the
(fast) Fourier transform, \(\sqrt{P}\) the square-root of the
power-spectrum in harmonic space, and \(\xi\) standard Gaussian random
variables. In the implementation in \texttt{NIFTy.re} and
\texttt{NIFTy}, the user can choose between two adaptive kernel models,
a non-parametric kernel \(\sqrt{P}\) and a Matérn kernel \(\sqrt{P}\)
(Arras et al., 2022; Guardiani et al., 2022 for details on their
implementation). A code example that initializes a non-parametric GP
prior for a \(128 \times 128\) space with unit volume is shown in the
following.

\begin{Shaded}
\begin{Highlighting}[]
\ImportTok{from}\NormalTok{ nifty8 }\ImportTok{import}\NormalTok{ re }\ImportTok{as}\NormalTok{ jft}

\NormalTok{dims }\OperatorTok{=}\NormalTok{ (}\DecValTok{128}\NormalTok{, }\DecValTok{128}\NormalTok{)}
\NormalTok{cfm }\OperatorTok{=}\NormalTok{ jft.CorrelatedFieldMaker(}\StringTok{"cf"}\NormalTok{)}
\NormalTok{cfm.set\_amplitude\_total\_offset(offset\_mean}\OperatorTok{=}\DecValTok{2}\NormalTok{, offset\_std}\OperatorTok{=}\NormalTok{(}\FloatTok{1e{-}1}\NormalTok{, }\FloatTok{3e{-}2}\NormalTok{))}
\CommentTok{\# Parameters for the kernel and the regular 2D Cartesian grid for which}
\CommentTok{\# it is defined}
\NormalTok{cfm.add\_fluctuations(}
\NormalTok{  dims,}
\NormalTok{  distances}\OperatorTok{=}\BuiltInTok{tuple}\NormalTok{(}\FloatTok{1.0} \OperatorTok{/}\NormalTok{ d }\ControlFlowTok{for}\NormalTok{ d }\KeywordTok{in}\NormalTok{ dims),}
\NormalTok{  fluctuations}\OperatorTok{=}\NormalTok{(}\FloatTok{1.0}\NormalTok{, }\FloatTok{5e{-}1}\NormalTok{),}
\NormalTok{  loglogavgslope}\OperatorTok{=}\NormalTok{(}\OperatorTok{{-}}\FloatTok{3.0}\NormalTok{, }\FloatTok{2e{-}1}\NormalTok{),}
\NormalTok{  flexibility}\OperatorTok{=}\NormalTok{(}\FloatTok{1e0}\NormalTok{, }\FloatTok{2e{-}1}\NormalTok{),}
\NormalTok{  asperity}\OperatorTok{=}\NormalTok{(}\FloatTok{5e{-}1}\NormalTok{, }\FloatTok{5e{-}2}\NormalTok{),}
\NormalTok{  prefix}\OperatorTok{=}\StringTok{"ax1"}\NormalTok{,}
\NormalTok{  non\_parametric\_kind}\OperatorTok{=}\StringTok{"power"}\NormalTok{,}
\NormalTok{)}
\CommentTok{\# Get the forward model for the GP prior}
\NormalTok{correlated\_field }\OperatorTok{=}\NormalTok{ cfm.finalize()}
\end{Highlighting}
\end{Shaded}

Not all problems are well described by regularly spaced pixels. For more
complicated pixel spacings, \texttt{NIFTy.re} features Iterative Charted
Refinement (Edenhofer et al., 2022), a GP model for arbitrarily deformed
spaces. This model exploits nearest neighbor relations on various
coarsenings of the discretized modeled space and runs very efficiently
on GPUs. For one-dimensional problems with arbitrarily spaced pixels,
\texttt{NIFTy.re} also implements multiple flavors of Gauss-Markov
processes.

\hypertarget{building-up-complex-models}{%
\subsection{Building Up Complex
Models}\label{building-up-complex-models}}

Models are rarely just a GP prior. Commonly, a model contains at least a
few non-linearities that transform the GP prior or combine it with other
random variables. For building more complex models, \texttt{NIFTy.re}
provides a \texttt{Model} class that offers a somewhat familiar
object-oriented design yet is fully JAX compatible and functional under
the hood. The following code shows how to build a slightly more complex
model using the objects from the previous example.

\begin{Shaded}
\begin{Highlighting}[]
\ImportTok{from}\NormalTok{ jax }\ImportTok{import}\NormalTok{ numpy }\ImportTok{as}\NormalTok{ jnp}


\KeywordTok{class}\NormalTok{ Forward(jft.Model):}
  \KeywordTok{def} \FunctionTok{\_\_init\_\_}\NormalTok{(}\VariableTok{self}\NormalTok{, correlated\_field):}
    \VariableTok{self}\NormalTok{.\_cf }\OperatorTok{=}\NormalTok{ correlated\_field}
    \CommentTok{\# Tracks a callable with which the model can be initialized. This}
    \CommentTok{\# is not strictly required, but comes in handy when building deep}
    \CommentTok{\# models. Note, the init method (short for "initialization" method)}
    \CommentTok{\# is not to be confused with the prior, which is always standard}
    \CommentTok{\# Gaussian.}
    \BuiltInTok{super}\NormalTok{().}\FunctionTok{\_\_init\_\_}\NormalTok{(init}\OperatorTok{=}\NormalTok{correlated\_field.init)}

  \KeywordTok{def} \FunctionTok{\_\_call\_\_}\NormalTok{(}\VariableTok{self}\NormalTok{, x):}
    \CommentTok{\# }\AlertTok{NOTE}\CommentTok{, any kind of masking of the output, non{-}linear and linear}
    \CommentTok{\# transformation could be carried out here. Models can also be}
    \CommentTok{\# combined and nested in any way and form.}
    \ControlFlowTok{return}\NormalTok{ jnp.exp(}\VariableTok{self}\NormalTok{.\_cf(x))}


\NormalTok{forward }\OperatorTok{=}\NormalTok{ Forward(correlated\_field)}

\NormalTok{data }\OperatorTok{=}\NormalTok{ jnp.load(}\StringTok{"data.npy"}\NormalTok{)}
\NormalTok{lh }\OperatorTok{=}\NormalTok{ jft.Poissonian(data).amend(forward)}
\end{Highlighting}
\end{Shaded}

All GP models in \texttt{NIFTy.re} as well as all likelihoods behave
like instances of \texttt{jft.Model}, meaning that JAX understands what
it means if a computation involves \texttt{self}, other
\texttt{jft.Model} instances, or their attributes. In other words,
\texttt{correlated\_field}, \texttt{forward}, and \texttt{lh} from the
code snippets shown here are all so-called pytrees in JAX, and, for
example, the following is valid code
\texttt{jax.jit(lambda\ l,\ x:\ l(x))(lh,\ x0)} with \texttt{x0} some
arbitrarily chosen valid input to \texttt{lh}. Inspired by equinox
(Kidger \& Garcia, 2021), individual attributes of the class can be
marked as non-static or static via
\texttt{dataclass.field(metadata=dict(static=...))} for the purpose of
compiling. Depending on the value, JAX will either treat the attribute
as an unknown placeholder or as a known concrete attribute and
potentially inline it during compilation. This mechanism is extensively
used in likelihoods to avoid inlining large constants such as the data
and to avoid expensive re-compilations whenever possible.

\hypertarget{variational-inference}{%
\subsection{Variational Inference}\label{variational-inference}}

\texttt{NIFTy.re} is built for models with millions to billions of
degrees of freedom. To probe the posterior efficiently and accurately,
\texttt{NIFTy.re} relies on VI. Specifically, \texttt{NIFTy.re}
implements Metric Gaussian Variational Inference (MGVI) and its
successor geometric Variational Inference (geoVI) (Frank et al., 2021;
Frank, 2022; Knollmüller \& Enßlin, 2019). At the core of both MGVI and
geoVI lies an alternating procedure in which one switches between
optimizing the Kullback--Leibler divergence for a specific shape of the
variational posterior and updating the shape of the variational
posterior. MGVI and geoVI define the variational posterior via samples,
specifically, via samples drawn around an expansion point. The samples
in MGVI and geoVI exploit model-intrinsic knowledge of the posterior's
approximate shape, encoded in the Fisher information metric and the
prior curvature (Frank et al., 2021).

\texttt{NIFTy.re} allows for much finer control over the way samples are
drawn and updated compared to \texttt{NIFTy}. \texttt{NIFTy.re} exposes
stand-alone functions for drawing MGVI and geoVI samples from any
arbitrary model with a likelihood from \texttt{NIFTy.re} and a forward
model that is differentiable by JAX. In addition to stand-alone sampling
functions, \texttt{NIFTy.re} provides tools to configure and execute the
alternating Kullback--Leibler divergence optimization and sample
adaption at a lower abstraction level. These tools are provided in a
JAXopt/Optax-style optimizer class (Blondel et al., 2022; DeepMind et
al., 2020).

A typical minimization with \texttt{NIFTy.re} is shown in the following.
It retrieves six independent, antithetically mirrored samples from the
approximate posterior via 25 iterations of alternating between
optimization and sample adaption. The final result is stored in the
\texttt{samples} variable. A convenient one-shot wrapper for the code
below is \texttt{jft.optimize\_kl}. By virtue of all modeling tools in
\texttt{NIFTy.re} being written in JAX, it is also possible to combine
\texttt{NIFTy.re} tools with BlackJAX (Cabezas \& Louf, 2023) or any
other posterior sampler in the JAX ecosystem.

\begin{Shaded}
\begin{Highlighting}[]
\ImportTok{from}\NormalTok{ jax }\ImportTok{import}\NormalTok{ random}

\NormalTok{key }\OperatorTok{=}\NormalTok{ random.PRNGKey(}\DecValTok{42}\NormalTok{)}
\NormalTok{key, sk }\OperatorTok{=}\NormalTok{ random.split(key, }\DecValTok{2}\NormalTok{)}
\CommentTok{\# NIFTy is agnostic w.r.t. the type of inputs it gets as long as they}
\CommentTok{\# support core arithmetic properties. Tell NIFTy to treat our parameter}
\CommentTok{\# dictionary as a vector.}
\NormalTok{samples }\OperatorTok{=}\NormalTok{ jft.Samples(pos}\OperatorTok{=}\NormalTok{jft.Vector(lh.init(sk)), samples}\OperatorTok{=}\VariableTok{None}\NormalTok{)}

\NormalTok{delta }\OperatorTok{=} \FloatTok{1e{-}4}
\NormalTok{absdelta }\OperatorTok{=}\NormalTok{ delta }\OperatorTok{*}\NormalTok{ jft.size(samples.pos)}

\NormalTok{opt\_vi }\OperatorTok{=}\NormalTok{ jft.OptimizeVI(lh, n\_total\_iterations}\OperatorTok{=}\DecValTok{25}\NormalTok{)}
\NormalTok{opt\_vi\_st }\OperatorTok{=}\NormalTok{ opt\_vi.init\_state(}
\NormalTok{  key,}
  \CommentTok{\# Implicit definition for the accuracy of the KL{-}divergence}
  \CommentTok{\# approximation; typically on the order of 2{-}12}
\NormalTok{  n\_samples}\OperatorTok{=}\KeywordTok{lambda}\NormalTok{ i: }\DecValTok{1} \ControlFlowTok{if}\NormalTok{ i }\OperatorTok{\textless{}} \DecValTok{2} \ControlFlowTok{else}\NormalTok{ (}\DecValTok{2} \ControlFlowTok{if}\NormalTok{ i }\OperatorTok{\textless{}} \DecValTok{4} \ControlFlowTok{else} \DecValTok{6}\NormalTok{),}
  \CommentTok{\# Parametrize the conjugate gradient method at the heart of the}
  \CommentTok{\# sample{-}drawing}
\NormalTok{  draw\_linear\_kwargs}\OperatorTok{=}\BuiltInTok{dict}\NormalTok{(}
\NormalTok{    cg\_name}\OperatorTok{=}\StringTok{"SL"}\NormalTok{, cg\_kwargs}\OperatorTok{=}\BuiltInTok{dict}\NormalTok{(absdelta}\OperatorTok{=}\NormalTok{absdelta }\OperatorTok{/} \FloatTok{10.0}\NormalTok{, maxiter}\OperatorTok{=}\DecValTok{100}\NormalTok{)}
\NormalTok{  ),}
  \CommentTok{\# Parametrize the minimizer in the nonlinear update of the samples}
\NormalTok{  nonlinearly\_update\_kwargs}\OperatorTok{=}\BuiltInTok{dict}\NormalTok{(}
\NormalTok{    minimize\_kwargs}\OperatorTok{=}\BuiltInTok{dict}\NormalTok{(}
\NormalTok{      name}\OperatorTok{=}\StringTok{"SN"}\NormalTok{, xtol}\OperatorTok{=}\NormalTok{delta, cg\_kwargs}\OperatorTok{=}\BuiltInTok{dict}\NormalTok{(name}\OperatorTok{=}\VariableTok{None}\NormalTok{), maxiter}\OperatorTok{=}\DecValTok{5}
\NormalTok{    )}
\NormalTok{  ),}
  \CommentTok{\# Parametrize the minimization of the KL{-}divergence cost potential}
\NormalTok{  kl\_kwargs}\OperatorTok{=}\BuiltInTok{dict}\NormalTok{(minimize\_kwargs}\OperatorTok{=}\BuiltInTok{dict}\NormalTok{(name}\OperatorTok{=}\StringTok{"M"}\NormalTok{, xtol}\OperatorTok{=}\NormalTok{delta, maxiter}\OperatorTok{=}\DecValTok{35}\NormalTok{)),}
\NormalTok{  sample\_mode}\OperatorTok{=}\StringTok{"nonlinear\_resample"}\NormalTok{,}
\NormalTok{)}
\ControlFlowTok{for}\NormalTok{ i }\KeywordTok{in} \BuiltInTok{range}\NormalTok{(opt\_vi.n\_total\_iterations):}
  \BuiltInTok{print}\NormalTok{(}\SpecialStringTok{f"Iteration }\SpecialCharTok{\{i}\OperatorTok{+}\DecValTok{1}\SpecialCharTok{:04d\}}\SpecialStringTok{"}\NormalTok{)}
  \CommentTok{\# Continuously update the samples of the approximate posterior}
  \CommentTok{\# distribution}
\NormalTok{  samples, opt\_vi\_st }\OperatorTok{=}\NormalTok{ opt\_vi.update(samples, opt\_vi\_st)}
  \BuiltInTok{print}\NormalTok{(opt\_vi.get\_status\_message(samples, opt\_vi\_st))}
\end{Highlighting}
\end{Shaded}

\begin{figure}
\centering
\includegraphics{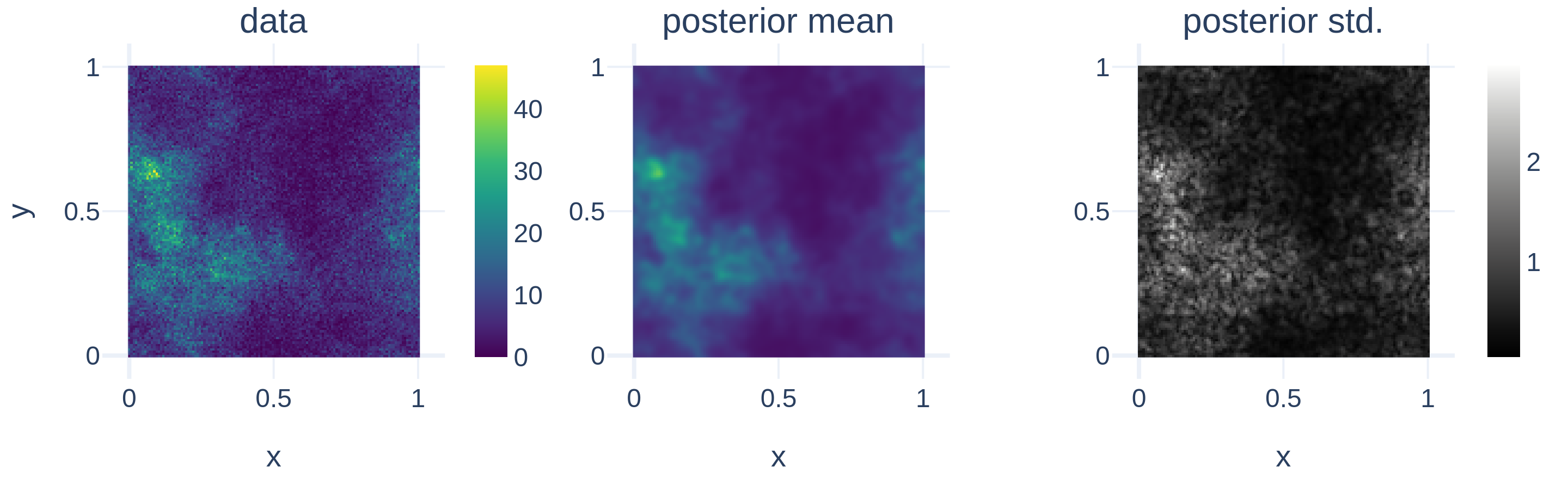}
\caption{Data (left), posterior mean (middle), and posterior uncertainty
(right) for a simple toy
example.\label{fig:minimal_reconstruction_data_mean_std}}
\end{figure}

\autoref{fig:minimal_reconstruction_data_mean_std} shows an exemplary
posterior reconstruction employing the above model. The posterior mean
agrees with the data but removes noisy structures. The posterior
standard deviation is approximately equal to typical differences between
the posterior mean and the data.

\hypertarget{performance-of-compared-to}{%
\subsection{\texorpdfstring{Performance of \texttt{NIFTy.re} compared to
\texttt{NIFTy}}{Performance of  compared to }}\label{performance-of-compared-to}}

We test the performance of \texttt{NIFTy.re} against \texttt{NIFTy} for
the simple yet representative model from above. To assess the
performance, we compare the time required to apply
\(M_p := F_p + \mathds{1}\) to random input with \(F_p\) denoting the
Fisher metric of the overall likelihood at position \(p\) and
\(\mathds{1}\) the identity matrix. Within \texttt{NIFTy.re}, the Fisher
metric of the overall likelihood is decomposed into
\(J_{f,p}^\dagger N^{-1} J_{f,p}\) with \(J_{f,p}\) the implicit
Jacobian of the forward model \(f\) at \(p\) and \(N^{-1}\) the
Fisher-metric of the Poisson likelihood. We choose to benchmark \(M_p\)
as a typical VI minimization in \texttt{NIFTy.re} and \texttt{NIFTy} is
dominated by calls to this function.

\begin{figure}
\centering
\includegraphics{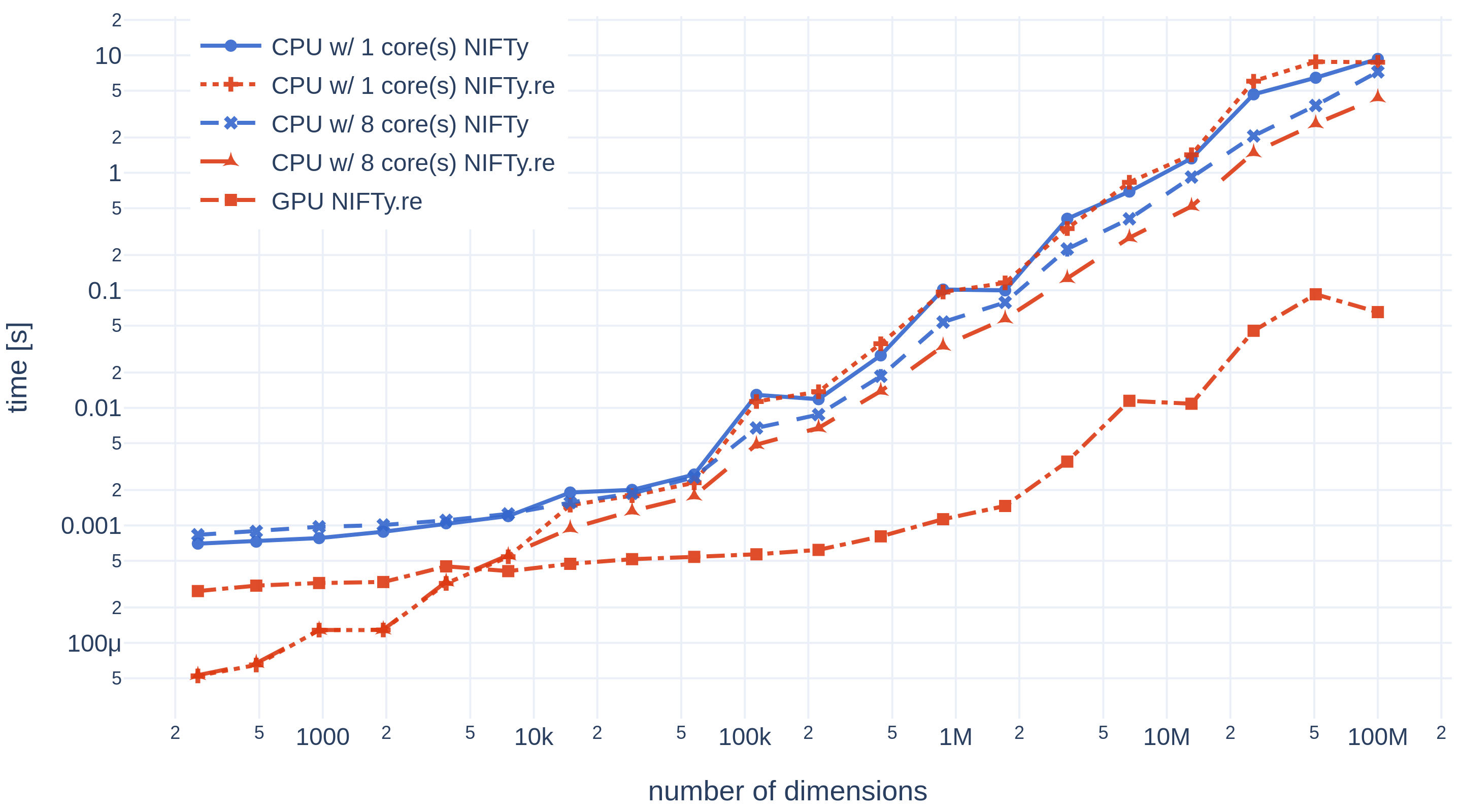}
\caption{Median evaluation time of applying the Fisher metric plus the
identity metric to random input for \texttt{NIFTy.re} and \texttt{NIFTy}
on the CPU (one and eight core(s) of an Intel Xeon Platinum 8358 CPU
clocked at 2.60G Hz) and the GPU (A100 SXM4 80 GB HBM2). The quantile
range from the 16\%- to the 84\%-quantile is obscured by the marker
symbols.\label{fig:benchmark_nthreads=1+8_devices=cpu+gpu}}
\end{figure}

\autoref{fig:benchmark_nthreads=1+8_devices=cpu+gpu} shows the median
evaluation time in \texttt{NIFTy} of applying \(M_p\) to new, random
tangent positions and the evaluation time in \texttt{NIFTy.re} of
building \(M_p\) and applying it to new, random tangent positions for
exponentially larger models. The 16\%-quantiles and the 84\%-quantiles
of the timings are obscured by the marker symbols. We chose to exclude
the build time of \(M_p\) in \texttt{NIFTy} from the comparison, putting
\texttt{NIFTy} at an advantage, as its automatic differentiation is
built around calls to \(M_p\) with \(p\) rarely varying. We ran the
benchmark on one CPU core, eight CPU cores, and on a GPU on a
compute-node with an Intel Xeon Platinum 8358 CPU clocked at 2.60G Hz
and an NVIDIA A100 SXM4 80 GB HBM2 GPU. The benchmark used
\texttt{jax==0.4.23} and \texttt{jaxlib==0.4.23+cuda12.cudnn89}. We vary
the size of the model by increasing the size of the two-dimensional
square image grid.

For small image sizes, \texttt{NIFTy.re} on the CPU is about one order
of magnitude faster than \texttt{NIFTy}. Both reach about the same
performance at an image size of roughly 15,000 pixels and continue to
perform roughly the same for larger image sizes. The performance
increases by a factor of three to four with eight cores for
\texttt{NIFTy.re} and \texttt{NIFTy}, although \texttt{NIFTy.re} is
slightly better at using the additional cores. On the GPU,
\texttt{NIFTy.re} is consistently about one to two orders of magnitude
faster than \texttt{NIFTy} for images larger than 100,000 pixels.

We believe the performance benefits of \texttt{NIFTy.re} on the CPU for
small models stem from the reduced Python overhead by just-in-time
compiling computations. At image sizes larger than roughly 15,000
pixels, both evaluation times are dominated by the fast Fourier
transform and are hence roughly the same as both use the same underlying
implementation (Reinecke, 2024). Models in \texttt{NIFTy.re} and
\texttt{NIFTy} are often well aligned with GPU programming models and
thus consistently perform well on the GPU. Modeling components such as
the new GP models implemented in \texttt{NIFTy.re} are even better
aligned with GPU programming paradigms and yield even higher performance
gains (Edenhofer et al., 2022).

\hypertarget{conclusion}{%
\section{Conclusion}\label{conclusion}}

\texttt{NIFTy.re} implements the core GP and VI machinery of the
Bayesian imaging package \texttt{NIFTy} in JAX. The rewrite moves much
of the heavy-lifting from home-grown solutions to JAX, and we envision
significant gains in maintainability of \texttt{NIFTy.re} and a faster
development cycle moving forward. The rewrite accelerates typical models
written in \texttt{NIFTy} by one to two orders of magnitude, lays the
foundation for new types of inference machineries by enabling higher
order derivatives via JAX, and enables the interoperability of
\texttt{NIFTy}'s VI and GP methods with the JAX machine learning
ecosystem.

\hypertarget{acknowledgements}{%
\section{Acknowledgements}\label{acknowledgements}}

Gordian Edenhofer acknowledges support from the German Academic
Scholarship Foundation in the form of a PhD scholarship
(``Promotionsstipendium der Studienstiftung des Deutschen Volkes'').
Philipp Frank acknowledges funding through the German Federal Ministry
of Education and Research for the project ``ErUM-IFT:
Informationsfeldtheorie für Experimente an Großforschungsanlagen''
(Förderkennzeichen: 05D23EO1). Jakob Roth acknowledges financial support
by the German Federal Ministry of Education and Research (BMBF) under
grant 05A20W01 (Verbundprojekt D-MeerKAT). Matteo Guardiani, Vincent
Eberle, and Margret Westerkamp acknowledge financial support from the
``Deutsches Zentrum für Luft- und Raumfahrt e.V.'' (DLR) through the
project Universal Bayesian Imaging Kit (UBIK, Förderkennzeichen
50OO2103). Lukas Scheel-Platz acknowledges funding from the European
Research Council (ERC) under the European Union's Horizon Europe
research and innovation programme under grant agreement No 101041936
(EchoLux).

\hypertarget{references}{%
\section{References}\label{references}}

\hypertarget{refs}{}
\begin{CSLReferences}{1}{0}
\leavevmode\hypertarget{ref-Sivaram2015}{}%
Ambikasaran, S., Foreman-Mackey, D., Greengard, L., Hogg, D. W., \&
O'Neil, M. (2015). {Fast Direct Methods for Gaussian Processes}.
\emph{IEEE Transactions on Pattern Analysis and Machine Intelligence},
\emph{38}, 252. \url{https://doi.org/10.1109/TPAMI.2015.2448083}

\leavevmode\hypertarget{ref-Arras2019NIFTy}{}%
Arras, P., Baltac, M., Ensslin, T. A., Frank, P., Hutschenreuter, S.,
Knollmueller, J., Leike, R., Newrzella, M.-N., Platz, L., Reinecke, M.,
\& Stadler, J. (2019). \emph{{NIFTy5: Numerical Information Field Theory
v5}}. Astrophysics Source Code Library, record ascl:1903.008.

\leavevmode\hypertarget{ref-Arras2022}{}%
Arras, P., Frank, P., Haim, P., Knollmüller, J., Leike, R. H., Reinecke,
M., \& Enßlin, T. A. (2022). {Variable structures in M87* from space,
time and frequency resolved interferometry}. \emph{Nature Astronomy},
\emph{6}, 259--269. \url{https://doi.org/10.1038/s41550-021-01548-0}

\leavevmode\hypertarget{ref-Arras2019}{}%
Arras, P., Frank, P., Leike, R., Westermann, R., \& Enßlin, T. A.
(2019). {Unified radio interferometric calibration and imaging with
joint uncertainty quantification}. \emph{Astronomy \& Astrophysics},
\emph{627}, A134. \url{https://doi.org/10.1051/0004-6361/201935555}

\leavevmode\hypertarget{ref-Bingham2019}{}%
Bingham, E., Chen, J. P., Jankowiak, M., Obermeyer, F., Pradhan, N.,
Karaletsos, T., Singh, R., Szerlip, P. A., Horsfall, P., \& Goodman, N.
D. (2019). Pyro: Deep universal probabilistic programming. \emph{Journal
of Machine Learning Research}, \emph{20}, 28:1--28:6.
\url{http://jmlr.org/papers/v20/18-403.html}

\leavevmode\hypertarget{ref-Blondel2021}{}%
Blondel, M., Berthet, Q., Cuturi, M., Frostig, R., Hoyer, S.,
Llinares-Lopez, F., Pedregosa, F., \& Vert, J.-P. (2022).
\emph{Efficient and modular implicit differentiation}. \emph{35},
5230--5242.
\url{https://proceedings.neurips.cc/paper_files/paper/2022/file/228b9279ecf9bbafe582406850c57115-Paper-Conference.pdf}

\leavevmode\hypertarget{ref-Jax2018}{}%
Bradbury, J., Frostig, R., Hawkins, P., Johnson, M. J., Leary, C.,
Maclaurin, D., Necula, G., Paszke, A., VanderPlas, J., Wanderman-Milne,
S., \& Zhang, Q. (2018). \emph{{JAX}: Composable transformations of
{P}ython+{N}um{P}y programs} (Version 0.3.13) {[}Computer software{]}.
\url{http://github.com/google/jax}

\leavevmode\hypertarget{ref-blackjax2020}{}%
Cabezas, L., Alberto, \& Louf, R. (2023). \emph{{B}lackjax: A sampling
library for {JAX}} (Version v1.1.0) {[}Computer software{]}.
\url{http://github.com/blackjax-devs/blackjax}

\leavevmode\hypertarget{ref-Carpenter2017}{}%
Carpenter, B., Gelman, A., Hoffman, M. D., Lee, D., Goodrich, B.,
Betancourt, M., Brubaker, M., Guo, J., Li, P., \& Riddell, A. (2017).
Stan: A probabilistic programming language. \emph{Journal of Statistical
Software}, \emph{76}(1), 1--32.
\url{https://doi.org/10.18637/jss.v076.i01}

\leavevmode\hypertarget{ref-Matthews2017}{}%
De G. Matthews, A. G., Van Der Wilk, M., Nickson, T., Fujii, K.,
Boukouvalas, A., León-Villagrá, P., Ghahramani, Z., \& Hensman, J.
(2017). GPflow: A gaussian process library using tensorflow.
\emph{Journal of Machine Learning Research}, \emph{18}(1), 1299--1304.

\leavevmode\hypertarget{ref-Deepmind2020Optax}{}%
DeepMind, Babuschkin, I., Baumli, K., Bell, A., Bhupatiraju, S., Bruce,
J., Buchlovsky, P., Budden, D., Cai, T., Clark, A., Danihelka, I.,
Dedieu, A., Fantacci, C., Godwin, J., Jones, C., Hemsley, R., Hennigan,
T., Hessel, M., Hou, S., \ldots{} Viola, F. (2020). \emph{The
{D}eep{M}ind {JAX} {E}cosystem}. \url{http://github.com/google-deepmind}

\leavevmode\hypertarget{ref-Eberle2023ButterflyImaging}{}%
Eberle, V., Frank, P., Stadler, J., Streit, S., \& Enßlin, T. (2023).
{Butterfly Transforms for Efficient Representation of Spatially Variant
Point Spread Functions in Bayesian Imaging}. \emph{Entropy},
\emph{25}(4), 652. \url{https://doi.org/10.3390/e25040652}

\leavevmode\hypertarget{ref-Eberle2023ButterflyImagingAlgorithm}{}%
Eberle, V., Frank, P., Stadler, J., Streit, S., \& Enßlin, T. (2022).
Efficient representations of spatially variant point spread functions
with butterfly transforms in bayesian imaging algorithms. \emph{Physical
Sciences Forum}, \emph{5}(1).
\url{https://doi.org/10.3390/psf2022005033}

\leavevmode\hypertarget{ref-Edenhofer2022}{}%
Edenhofer, G., Leike, R. H., Frank, P., \& Enßlin, T. A. (2022).
\emph{Sparse kernel gaussian processes through iterative charted
refinement (ICR)}. arXiv.
\url{https://doi.org/10.48550/ARXIV.2206.10634}

\leavevmode\hypertarget{ref-Edenhofer2023}{}%
Edenhofer, G., Zucker, C., Frank, P., Saydjari, A. K., Speagle, J. S.,
Finkbeiner, D., \& Enßlin, T. A. (2024). {A parsec-scale Galactic 3D
dust map out to 1.25 kpc from the Sun}. \emph{Astronomy \&
Astrophysics}, \emph{685}, A82.
\url{https://doi.org/10.1051/0004-6361/202347628}

\leavevmode\hypertarget{ref-ForemanMackey2013}{}%
Foreman-Mackey, D., Hogg, D. W., Lang, D., \& Goodman, J. (2013).
{emcee: The MCMC Hammer}. \emph{Publications of the Astronomical Society
of the Pacific}, \emph{125}(925), 306.
\url{https://doi.org/10.1086/670067}

\leavevmode\hypertarget{ref-ForemanMackey2024}{}%
Foreman-Mackey, D., Yu, W., Yadav, S., Becker, M. R., Caplar, N.,
Huppenkothen, D., Killestein, T., Tronsgaard, R., Rashid, T., \&
Schmerler, S. (2024). \emph{{dfm/tinygp: The tiniest of Gaussian Process
libraries}} (Version v0.3.0) {[}Computer software{]}. Zenodo.
\url{https://doi.org/10.5281/zenodo.10463641}

\leavevmode\hypertarget{ref-Frank2022}{}%
Frank, P. (2022). Geometric variational inference and its application to
bayesian imaging. \emph{Physical Sciences Forum}, \emph{5}(1).
\url{https://doi.org/10.3390/psf2022005006}

\leavevmode\hypertarget{ref-Frank2021}{}%
Frank, P., Leike, R. H., \& Enßlin, T. A. (2021). Geometric variational
inference. \emph{Entropy}, \emph{23}(7), 853.
\url{https://doi.org/10.3390/e23070853}

\leavevmode\hypertarget{ref-Frank2017}{}%
Frank, P., Steininger, T., \& Enßlin, T. A. (2017). {Field dynamics
inference via spectral density estimation}. \emph{Physical Review E},
\emph{96}(5), 052104. \url{https://doi.org/10.1103/PhysRevE.96.052104}

\leavevmode\hypertarget{ref-Guardiani2022}{}%
Guardiani, M., Frank, P., Kostić, A., Edenhofer, G., Roth, J., Uhlmann,
B., \& Enßlin, T. (2022). Causal, bayesian, \& non-parametric modeling
of the SARS-CoV-2 viral load distribution vs. Patient's age. \emph{PLOS
ONE}, \emph{17}(10), 1--21.
\url{https://doi.org/10.1371/journal.pone.0275011}

\leavevmode\hypertarget{ref-Hensman2015}{}%
Hensman, J., G. Matthews, A. G. de, \& Ghahramani, Z. (2015). Scalable
variational gaussian process classification. In G. Lebanon \& S. V. N.
Vishwanathan (Eds.), \emph{AISTATS} (Vol. 38). JMLR.org.
\url{http://dblp.uni-trier.de/db/conf/aistats/aistats2015.html\#HensmanMG15}

\leavevmode\hypertarget{ref-Hutschenreuter2022}{}%
Hutschenreuter, S., Anderson, C. S., Betti, S., Bower, G. C., Brown,
J.-A., Brüggen, M., Carretti, E., Clarke, T., Clegg, A., Costa, A.,
Croft, S., Van Eck, C., Gaensler, B. M., de Gasperin, F., Haverkorn, M.,
Heald, G., Hull, C. L. H., Inoue, M., Johnston-Hollitt, M., \ldots{}
Enßlin, T. A. (2022). {The Galactic Faraday rotation sky 2020}.
\emph{Astronomy \& Astrophysics}, \emph{657}, A43.
\url{https://doi.org/10.1051/0004-6361/202140486}

\leavevmode\hypertarget{ref-Hutschenreuter2023}{}%
Hutschenreuter, Sebastian, Haverkorn, M., Frank, P., Raycheva, N. C., \&
Enßlin, T. A. (2023). {Disentangling the Faraday rotation sky}.
\emph{arXiv e-Prints}, arXiv:2304.12350.
\url{https://doi.org/10.48550/arXiv.2304.12350}

\leavevmode\hypertarget{ref-Kidger2021}{}%
Kidger, P., \& Garcia, C. (2021). {E}quinox: Neural networks in {JAX}
via callable {P}y{T}rees and filtered transformations.
\emph{Differentiable Programming Workshop at Neural Information
Processing Systems 2021}.

\leavevmode\hypertarget{ref-Knollmueller2019}{}%
Knollmüller, J., \& Enßlin, T. A. (2019). \emph{Metric gaussian
variational inference}. arXiv.
\url{https://doi.org/10.48550/ARXIV.1901.11033}

\leavevmode\hypertarget{ref-Koposov2023}{}%
Koposov, S., Speagle, J., Barbary, K., Ashton, G., Bennett, E., Buchner,
J., Scheffler, C., Cook, B., Talbot, C., Guillochon, J., Cubillos, P.,
Ramos, A. A., Johnson, B., Lang, D., Ilya, Dartiailh, M., Nitz, A.,
McCluskey, A., \& Archibald, A. (2023). \emph{Joshspeagle/dynesty:
v2.1.3} (Version v2.1.3) {[}Computer software{]}. Zenodo.
\url{https://doi.org/10.5281/zenodo.8408702}

\leavevmode\hypertarget{ref-Leike2022}{}%
Leike, R. H., Edenhofer, G., Knollmüller, J., Alig, C., Frank, P., \&
Enßlin, T. A. (2022). {The Galactic 3D large-scale dust distribution via
Gaussian process regression on spherical coordinates}. \emph{arXiv
e-Prints}, arXiv:2204.11715.
\url{https://doi.org/10.48550/arXiv.2204.11715}

\leavevmode\hypertarget{ref-Leike2019}{}%
Leike, R. H., \& Enßlin, T. A. (2019). {Charting nearby dust clouds
using Gaia data only}. \emph{Astronomy \& Astrophysics}, \emph{631},
A32. \url{https://doi.org/10.1051/0004-6361/201935093}

\leavevmode\hypertarget{ref-Leike2020}{}%
Leike, R. H., Glatzle, M., \& Enßlin, T. A. (2020). {Resolving nearby
dust clouds}. \emph{Astronomy \& Astrophysics}, \emph{639}, A138.
\url{https://doi.org/10.1051/0004-6361/202038169}

\leavevmode\hypertarget{ref-Mertsch2023}{}%
Mertsch, P., \& Phan, V. H. M. (2023). {Bayesian inference of
three-dimensional gas maps. II. Galactic HI}. \emph{Astronomy \&
Astrophysics}, \emph{671}, A54.
\url{https://doi.org/10.1051/0004-6361/202243326}

\leavevmode\hypertarget{ref-Phan2019}{}%
Phan, D., Pradhan, N., \& Jankowiak, M. (2019). {Composable Effects for
Flexible and Accelerated Probabilistic Programming in NumPyro}.
\emph{arXiv e-Prints}, arXiv:1912.11554.
\url{https://doi.org/10.48550/arXiv.1912.11554}

\leavevmode\hypertarget{ref-ducc0}{}%
Reinecke, M. (2024). \emph{{DUCC}: Distinctly useful code collection}
(Version 0.33.0) {[}Computer software{]}.
\url{https://gitlab.mpcdf.mpg.de/mtr/ducc}

\leavevmode\hypertarget{ref-Rezende2015}{}%
Rezende, D. J., \& Mohamed, S. (2015). Variational inference with
normalizing flows. \emph{Proceedings of the 32nd International
Conference on International Conference on Machine Learning - Volume 37},
1530--1538. \url{http://proceedings.mlr.press/v37/rezende15.html}

\leavevmode\hypertarget{ref-Roth2023DirectionDependentCalibration}{}%
Roth, Jakob, Arras, P., Reinecke, M., Perley, R. A., Westermann, R., \&
Enßlin, T. A. (2023). {Bayesian radio interferometric imaging with
direction-dependent calibration}. \emph{Astronomy \& Astrophysics},
\emph{678}, A177. \url{https://doi.org/10.1051/0004-6361/202346851}

\leavevmode\hypertarget{ref-Roth2023FastCadenceHighContrastImaging}{}%
Roth, J., Li Causi, G., Testa, V., Arras, P., \& Ensslin, T. A. (2023).
{Fast-cadence High-contrast Imaging with Information Field Theory}.
\emph{The Astronomical Journal}, \emph{165}(3), 86.
\url{https://doi.org/10.3847/1538-3881/acabc1}

\leavevmode\hypertarget{ref-Salvatier2016}{}%
Salvatier, J., Wiecki, T. V., \& Fonnesbeck, C. (2016). Probabilistic
programming in python using PyMC3. \emph{PeerJ Computer Science},
\emph{2}, e55. \url{https://doi.org/10.7717/peerj-cs.55}

\leavevmode\hypertarget{ref-ScheelPlatz2023}{}%
Scheel-Platz, L. I., Knollmüller, J., Arras, P., Frank, P., Reinecke,
M., Jüstel, D., \& Enßlin, T. A. (2023). {Multicomponent imaging of the
Fermi gamma-ray sky in the spatio-spectral domain}. \emph{Astronomy \&
Astrophysics}, \emph{680}, A2.
\url{https://doi.org/10.1051/0004-6361/202243819}

\leavevmode\hypertarget{ref-Selig2013}{}%
Selig, M., Bell, M. R., Junklewitz, H., Oppermann, N., Reinecke, M.,
Greiner, M., Pachajoa, C., \& Enßlin, T. A. (2013). {NIFTY - Numerical
Information Field Theory. A versatile PYTHON library for signal
inference}. In \emph{Astronomy \& Astrophysics} (Vol. 554, p. A26).
\url{https://doi.org/10.1051/0004-6361/201321236}

\leavevmode\hypertarget{ref-Speagle2020}{}%
Speagle, J. S. (2020). {DYNESTY: a dynamic nested sampling package for
estimating Bayesian posteriors and evidences}. \emph{Monthly Notices of
the RAS}, \emph{493}(3), 3132--3158.
\url{https://doi.org/10.1093/mnras/staa278}

\leavevmode\hypertarget{ref-Steiniger2017}{}%
Steininger, T., Dixit, J., Frank, P., Greiner, M., Hutschenreuter, S.,
Knollmüller, J., Leike, R. H., Porqueres, N., Pumpe, D., Reinecke, M.,
Šraml, M., Varady, C., \& Enßlin, T. A. (2019). {NIFTy 3 - Numerical
Information Field Theory: A Python Framework for Multicomponent Signal
Inference on HPC Clusters}. \emph{Annalen Der Physik}, \emph{531}(3),
1800290. \url{https://doi.org/10.1002/andp.201800290}

\leavevmode\hypertarget{ref-Tsouros2023}{}%
Tsouros, A., Edenhofer, G., Enßlin, T., Mastorakis, M., \& Pavlidou, V.
(2024). {Reconstructing Galactic magnetic fields from local measurements
for backtracking ultra-high-energy cosmic rays}. \emph{Astronomy \&
Astrophysics}, \emph{681}, A111.
\url{https://doi.org/10.1051/0004-6361/202346423}

\leavevmode\hypertarget{ref-Welling2021}{}%
Welling, C., Frank, P., Enßlin, T., \& Nelles, A. (2021).
{Reconstructing non-repeating radio pulses with Information Field
Theory}. \emph{Journal of Cosmology and Astroparticle Physics},
\emph{2021}(4), 071. \url{https://doi.org/10.1088/1475-7516/2021/04/071}

\leavevmode\hypertarget{ref-Westerkamp2023}{}%
Westerkamp, M., Eberle, V., Guardiani, M., Frank, P., Platz, L., Arras,
P., Knollmüller, J., Stadler, J., \& Enßlin, T. (2023). {First
spatio-spectral Bayesian imaging of SN1006 in X-ray}. \emph{arXiv
e-Prints}, arXiv:2308.09176.
\url{https://doi.org/10.48550/arXiv.2308.09176}

\end{CSLReferences}

\end{document}